\documentclass[twocolumn]{aastex63}

\usepackage{amssymb}
\usepackage{multirow}

\usepackage{xcolor}

\newcommand{\report}[1]{{\textcolor{black}{#1}}}

\usepackage{changepage}

\received{MM DD, 2020}
\revised{MM DD, 2020}
\accepted{13 June 2020}
\submitjournal{ApJL}

\shorttitle{Image based classification of variable stars}
\shortauthors{Szklen\'ar et al.}

\begin{document}

\title{Image-based classification of variable stars - First results on OGLE data}

\author[0000-0002-5610-7697]{T. Szklen\'ar}
\email{szklenar.tamas@csfk.mta.hu}

\affiliation{Konkoly Observatory, Research Centre for Astronomy and Earth Sciences,\\ H-1121 Budapest, Konkoly Thege Mikl\'os \'ut 15-17, Hungary\\}
\affiliation{MTA CSFK Lend\"ulet Near-Field Cosmology Research Group\\}

\author[0000-0002-8585-4544]{A. B\'odi}
\affiliation{Konkoly Observatory, Research Centre for Astronomy and Earth Sciences,\\ H-1121 Budapest, Konkoly Thege Mikl\'os \'ut 15-17, Hungary\\}
\affiliation{MTA CSFK Lend\"ulet Near-Field Cosmology Research Group\\}
\affiliation{ELTE E\"otv\"os Lor\'and University, Institute of Physics, Budapest, Hungary\\}

\author[0000-0003-3759-7616]{D. Tarczay-Neh\'ez}
\affiliation{Konkoly Observatory, Research Centre for Astronomy and Earth Sciences,\\ H-1121 Budapest, Konkoly Thege Mikl\'os \'ut 15-17, Hungary\\}
\affiliation{MTA CSFK Lend\"ulet Near-Field Cosmology Research Group\\}

\author[0000-0002-6471-8607]{K. Vida}
\affiliation{Konkoly Observatory, Research Centre for Astronomy and Earth Sciences,\\ H-1121 Budapest, Konkoly Thege Mikl\'os \'ut 15-17, Hungary\\}
\affiliation{MTA CSFK Lend\"ulet Near-Field Cosmology Research Group\\}
\affiliation{ELTE E\"otv\"os Lor\'and University, Institute of Physics, Budapest, Hungary\\}

\author[0000-0002-1326-1686]{G. Marton}
\affiliation{Konkoly Observatory, Research Centre for Astronomy and Earth Sciences,\\ H-1121 Budapest, Konkoly Thege Mikl\'os \'ut 15-17, Hungary\\}
\affiliation{ELTE E\"otv\"os Lor\'and University, Institute of Physics, Budapest, Hungary\\}

\author[0000-0002-0686-7479]{Gy. Mez\H o}
\affiliation{Konkoly Observatory, Research Centre for Astronomy and Earth Sciences,\\ H-1121 Budapest, Konkoly Thege Mikl\'os \'ut 15-17, Hungary\\}

\author[0000-0001-9394-3531]{A. Forr\'o}
\affiliation{Konkoly Observatory, Research Centre for Astronomy and Earth Sciences,\\ H-1121 Budapest, Konkoly Thege Mikl\'os \'ut 15-17, Hungary\\}
\affiliation{MTA CSFK Lend\"ulet Near-Field Cosmology Research Group\\}
\affiliation{E\"otv\"os Lor\'and University, P\'azm\'any P\'eter s\'et\'any 1/A,  Budapest, Hungary\\}

\author[0000-0002-3258-1909]{R. Szab\'o}
\affiliation{Konkoly Observatory, Research Centre for Astronomy and Earth Sciences,\\ H-1121 Budapest, Konkoly Thege Mikl\'os \'ut 15-17, Hungary\\}
\affiliation{MTA CSFK Lend\"ulet Near-Field Cosmology Research Group\\}
\affiliation{ELTE E\"otv\"os Lor\'and University, Institute of Physics, Budapest, Hungary\\}




\begin{abstract}
\report{Recently}, machine learning methods presented a viable solution for automated classification of image-based data in various research fields and business applications. Scientists require a fast and reliable solution to be able to handle the always growing enormous amount of data in astronomy. However, so far astronomers have been \report{mainly} classifying  variable star  light curves based on various pre-computed statistics and light curve parameters.
In this work we use \report{an} image-based Convolutional Neural Network to classify the different types of variable stars. We used images of phase-folded light curves from the OGLE-III \report{survey for training, validating and testing and used OGLE-IV survey as an independent data set for testing}.
\report{After the training phase, our neural network was able to classify the different types between 80 and 99\%, and 77--98\% accuracy for OGLE-III and OGLE-IV, respectively}.
\end{abstract}



\section{Introduction} 
\label{sec:intro}
Most recent space-borne (e.g. Kepler, see \citealp{KeplerMission};  Gaia, see \citealp{GaiaMission}; TESS, see \citealp{TESSMission}) and ground-based sky surveys (e.g. SDSS, see \citealp{SDSS}; and LSST, see \citealp{LSST}) provide huge amount of data, that leads to a new level of challenge in data processing. This enormous quantity of data need to be analysed with fast and effective automated computer programming techniques. As a consequence, several machine learning algorithms became popular in astronomy.

Automatic classification of variable stars using machine learning methods mostly uses photometric data sets where objects are represented by their light curves. The classical approach of variable star classification relies on carefully selected features of the light curves, such as statistical metrics (like mean, standard deviation, kurtosis, skewness; see e.g. \citealp{Nun2015}), Fourier-decomposition \citep{Kim2016} or color information \citep{Miller2015}. The classifiers can be trained on manually designed \citep{Pashchenko2018,Hosenie2019} or computer-selected features \citep{Becker2020,Johnston2020} using known type of variable stars. Another opportunity to classify light curves is to use non-labeled data, which is called unsupervised learning. This method clusters  similar objects into groups instead of labelling them one-by-one \citep{Mackenzie2016,Valenzuela2018}.

Image-based classification is now in our everyday life: we use it in our phones, social network applications, cars, etc. Convolutional Neural Networks \citep[CNNs,][]{CNN_basic_article} -- a class of deep neural networks -- can distinguish  humans, animals and various objects. If CNNs are well trained, they can learn very fine features of an image (e.g. face recognition), therefore this kind of technology is now widely used in many scientific fields, for example geology, biology or even in medicine to recognise tumours and other diseases in the human body \citep[e.g.][]{braintumor}. Recently, CNNs have been successfully applied to astronomical problems as well, like real/bogus separation \citep{Gieseke2017}, cold gas study in galaxies \citep{Dawson2020}, supernova classification \citep{Moller2020}, LIGO data classification \citep{George2018}, and exoplanet candidate classification \citep{Osborn2020}. \cite{Hon2018b} trained a convolutional network on 2D images of red giant power spectra to detect solar like oscillations, and later used the method to classify the evolutionary states of red giants observed by Kepler \citep{Hon2018a}. \citet{Carrasco-Davis2019} designed a recurrent convolutional neural network to classify astronomical objects using image sequences, however their approach does not compute the light curves itself.

\report{An approach similar to ours was used by \citet{2017_Deep_LearntCNN}, who transformed the raw light curves into \textit{dmdt} space and mapped the results to 2D images. These images were then classified using a CNN. Moreover, two other works also took advantage of neural networks to classify variables stars. \citet{2018Deep_multi_surveyCNN} used a recurrent NN, which was fed by the light curve measurements one by one as individual points. \citet{Reccur-Deep_multi_survey} calculated the difference between consecutive measurement times and magnitude values, and classified the resulted pair of one-dimensional vectors using a CNN. However, the automatic classification of variable stars, which is fully based on the photometric light curves that are represented as images has not been performed so far to our knowledge.}

In this work, we present the first results of an image-based classification of phase-folded light curves of periodic variable stars with our deep neural network architecture trained \report{and validated} on OGLE-III and tested on OGLE-III and \report{independently on} OGLE-IV databases \citep{Udalski2008,Udalski2015}.
\report{The goal of our work was to test whether we are able to classify the phase-folded light curve images, focusing only on the shape of the light curves and neglecting period information.} The idea is very similar to the way human perception works when a traditional astronomer visually evaluates a light curve, i.e. deciding based on distinctive features and patterns. In this study we demonstrate that a deep neural network trained with light curve images can effectively used for classification. 

The paper is structured as follows: In Section\,\ref{sec:data} we discuss our data selection and handling, in Section\,\ref{sec:methods} we present our neural network and data sampling, while in Sections\,\ref{sec:results} and \ref{sec:conclusion} we show and conclude our results.

\section{Data}
\label{sec:data}

The aim of our project was to provide an effective and reliable solution for classifying variable stars by means of an image-based classification technique. As a first step, we restrict ourselves to use only periodic variable stars, so that images of phase-folded light curves can be used. 
Therefore, we need a \report{data set} that is classified in a reliable way and contains enough observations to create well-sampled phase-folded light curves.

\begin{figure}
   \centering
   \includegraphics[width=\columnwidth]{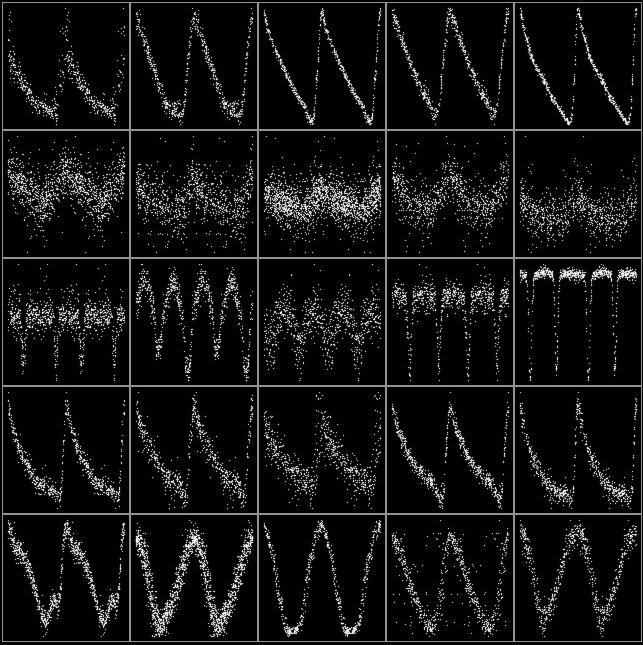}
      \caption{Gallery of phase-folded light curve images of different types of variables stars in OGLE-III. \report{The phases are in the [0..2] interval.} From top to bottom: ACep, DSct, ECL, RRLyr and T2Cep. \report{In case of pulsating variables the light curves are phased folded by their pulsation periods, while eclipsing binaries are folded by the orbital periods (i.e. twice the formal periods).} 
      }
      \label{fig:lcurves}
\end{figure}

\subsection{Observational data}

Many catalogs of variables stars are available from the literature. Among these the Optical Gravitational Lensing Experiment (OGLE; \citealt{Udalski2015}) provides one of the most extensive \report{data sets}, which have sufficient number of labels and labelled samples to train and test our neural networks. The survey is in its fourth phase, operating since 2010. 

OGLE observes the inner Galactic Bulge, the Magellanic Clouds and the Galactic disk. The observations are obtained in $V$ and $I$ bands, \report{ as the latter having about ten times more data points we chose to work with the $I$-band data only.}

The obtained light curves mostly have high signal-to-noise ratios and their types are confirmed by experts, which makes the sample very reliable. The OGLE-III catalog lists more than 450\,000 variable stars. Along the photometric data, the catalog includes basic parameters of the objects (such as coordinates, periods, mean magnitudes, amplitudes, parameters of the Fourier light curve decomposition), which can be used to our data preparation process.

The main variable star classes are divided into several sub-classes. However, in order to have homogeneous data, we only focused on 5 different main variable star types observed in the LMC field during OGLE-III. The chosen types were the following: Anomalous Cepheids \citep[ACep,][]{OGLEIII_ACep_T2Cep}, $\delta$ Scutis \citep[DSct,][]{OGLEIII_DSCT}, eclipsing binaries \citep[ECL,][]{OGLEIII_ECL}, RR Lyrae stars \citep[RRLyr,][]{OGLEIII_RRLyr}, and Type\,II Cepheids \citep[T2Cep,][]{OGLEIII_ACep_T2Cep}. The number of objects of each variable types is listed in Table\,\ref{Tab:Table1}.

\report{We converted the measured magnitudes of a given star into flux values with a zero point of 25, then normalized this data with the maximum brightness.}
Using the epochs and periods from the OGLE catalog, the light curves have been phase-folded and transformed into 8 bit images with a size of $128\times128$ pixels, with black background and white plotted dots (see Figure\,\ref{fig:lcurves}). \report{In case of pulsating variables we used the pulsation periods, while for eclipsing binaries we used the orbital periods (i.e. twice the formal periods) to phased-fold the light curves.}  Only the raw data were used, without sigma clipping and measurement error handling. In order to ensure that all of the representative light curve shapes are covered, the phased light curves are plotted in the [0..2] phase interval. These images served as the basis of our training sample.

One other purpose of our research was to know how well our trained model works with other observational data. This is why we generated light curves from the OGLE-IV database. Unfortunately,  $\delta$ Scuti stars have not been published yet, so we could use the following types only: Anomalous Cepheids \citep[ACep,][]{OGLEIV_ACEP}, Eclipsing binaries \citep[ECL,][]{OGLEIV_ECL}, RR Lyrae stars \citep[RRLyr,][]{OGLEIV_RRLYR}, and Type\,II Cepheids \citep[T2Cep,][]{OGLEIV_T2CEP}. The subtypes were not separated, we used the same method by the image generation.

\begin{figure*}
   \centering
   \includegraphics[width=\textwidth]{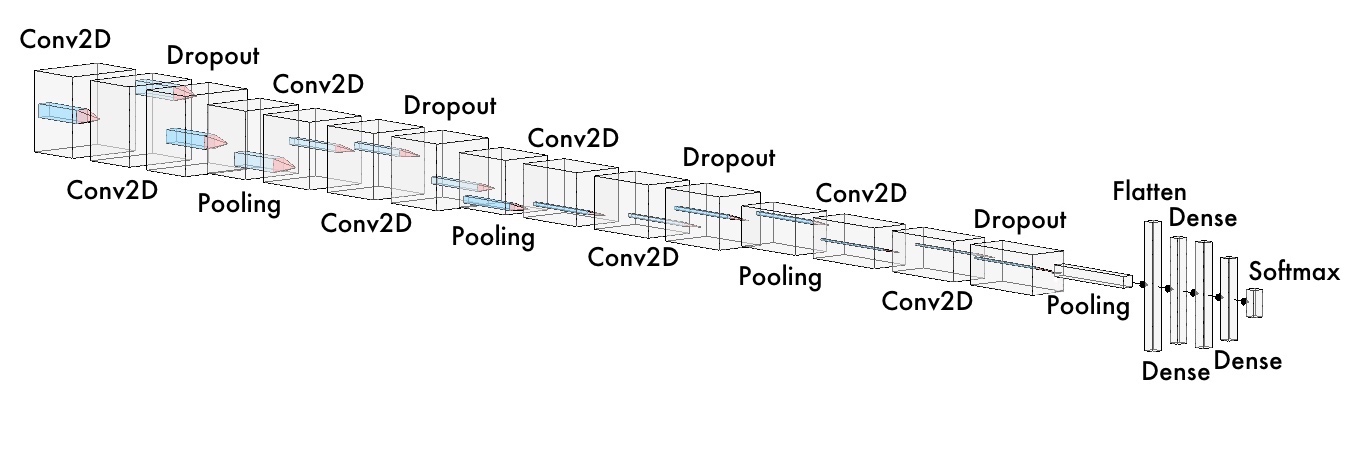}
      \caption{Schematic of the architecture of the designed CNN. }
    \label{fig:architecture}
\end{figure*}

\begin{table}
\caption{The number of variable stars in the original and in the augmented \report{data set}.}
\label{Tab:Table1}
\begin{tabular}{rcc}
\hline
\multicolumn{1}{c}{} & Non-augmented & Augmented \\
\hline
\hline
ACep                 & 83        & 25\,000  \\
DSct                 & 2\,696      & 25\,000  \\
ECL                  & 26\,121     & 25\,000  \\
RRLyr                & 24\,904     & 25\,000  \\
T2Cep                & 203       & 25\,000 \\
Total               & 54\,007 & 125\,000\\
\hline
\end{tabular}
\end{table}

\subsection{Data augmentation}
\label{sec:data_aug}
Highly unbalanced number of representatives in different classes, which is the case here (as it can seen in Table\,\ref{Tab:Table1}), may cause false machine learning output, therefore data augmentation was crucial in the pre-processing phase. The basic data augmentation methods usually use e.g. Gaussian noise, elastic transform, random brightness or contrast changes \cite[see e.g.][]{Shorten2019}; the images can also be mirrored or rotated. These methods allow us to create more duplicates, and it works well when classifying everyday objects: a mirrored cat is still a cat -- however, this is not true for light curves. 

To increase the sample of underrepresented classes, \report{randomly generated noise was sampled from a Gaussian distribution with zero mean and standard deviation equal to the given measured  error then added to the original light curves.} The augmented training data contained 125\,000 images, 25\,000 from each variable star type (the number of the eclipsing binaries were reduced).
\report{We took precautions to ensure that the training, testing and validating sets contain non-overlapping samples of the generated dummy light curves, i.e. for one given star the original light curve and its generated dummy multiples are used only in one of the aforementioned steps.}

\section{ Methods and models}
\label{sec:methods}

\subsection{DarkNet}
In order to investigate the effectiveness of a CNN on classifying the folded light curves, first we tested DarkNet \citep{DarkNet}, a GPU supported open-source software. We used a built-in, very simple convolutional neural network for the first training of our data. This was originally created for the CIFAR-10 \report{data set}%
\footnote{\url{https://www.cs.toronto.edu/~kriz/cifar.html}}%
, which is a test to classify images from 10 different classes of freely downloadable $28\times28$ pixel images of cars, dogs, cats, ships, etc.

Our first training package was not augmented, the classes had large differences in the amount of data. This set contained 54\,007 images, see Table\,\ref{Tab:Table1}.

Training a deep neural network requires vast amount of computational time and capacity.
To be able to test our first deep neural network on simple desktop computers, we created a much smaller image package. The first training was made with less than 500 images, but it took 1.5 hours to complete one training. We used a high-performance computer for this task, containing 4 Tesla V100 GPUs. \report{
Although the first results using the DarkNet framework were promising, due to the poor documentation, the complexity of architecting a network and the required time to preprocess data to match format with the DarkNet requirements, we decided to move to a more user-friendly framework.
}

\subsection{TensorFlow/Keras}
As the DarkNet package is poorly documented and not being maintained, we moved to compile a new neural network based on the TensorFlow/Keras framework. TensorFlow \citep{Tensorflow} is a free and open-source framework which is widely used in different machine learning applications. Keras \citep{Keras} is an open-source, high-level language and neural-network library for creating deep neural network with ease and it is officially supported in the TensorFlow core library since 2017. 
As a first step we recreated the previous CNN, now in Keras, using Python programming language. During the testing phase we used TensorBoard \citep{Tensorflow} to be able to visualize the differences, track the changes in training loss and accuracy.

\subsection{Our Convolutional Neural Network}
\label{sec:CNN_creation}
Convolutional networks use convolution instead of general matrix multiplication in their layers. A typical network architecture uses a mixture of convolutional, pooling and fully connected layers. Additionally, dropout layers can be added for regularization purposes. 
CNNs set the weights for the filter kernels during the learning process instead of using pre-set kernels as e.g. in early optical character recognition solutions: this independence from prior knowledge gives them great flexibility, and the ability to recognize features on different spatial scales in their consecutive layers. 

\report{
Figure\,\ref{fig:architecture} shows a schematic view of our CNN. Our model has a conventional structure, it consists of 2 convolutional, 1 dropout and 1 pooling layers in all 4 blocks. The resolution of input images is $128\times128$ pixels, the first two convolutional layers use a $16\times16$ pixels width convolutional window (known as ”kernel/filtersize”), with 1 pixel stride to run through the images. After this step, the second, third and fourth pair of convolutional layers use $8\times8$, $4\times4$ and $2\times2$ pixel-wide windows, respectively. 
This way, our model can learn the low-level features in the beginning of the training process and the high-level features during the last convolutional layers as well. All convolutional layers are using Rectified Linear Unit (ReLU) activation\footnote{f(x) = max(0,x)}. The output of the last convolution block is flattened and sent to a network of fully connected layers (“dense layers”). The last one is a softmax layer which is used to normalize the output and hence yields predictions (numbers between $0 - 1$) for all the 5 possible output labels.
The total number of trainable parameters were 1\,615\,685 altogether.
The tested hyperparameters and the final chosen ones are listed in Table\,\ref{Tab:hyperparameters}. }

\begin{table*}
\centering
\caption{\report{Hyperparameters of our Convolutional Neural Network}}
\label{Tab:hyperparameters}
\begin{tabular}{rcc}
\hline
Parameter & Tested values & Chosen value \\
\hline
\hline

\noalign{\smallskip}
\multicolumn{3}{c}{\emph{Architecture}} \\
\noalign{\smallskip}

Starting convolution window & [$8\times8$, $16\times16$ , $32\times32$] & $16\times16$ \\
Convolution stride & 1 & 1\\
Convolution padding & 0 & 0\\
Convolution activation & ReLU & ReLU \\
Dropout probability & [0.1--0.5] & 0.2 \\
Pooling type & MaxPooling & MaxPooling \\
Pooling size & [$2\times2$, $3\times3$] & $2\times2$ \\
Number of convolution layers & 8 & 8 \\
Number of pooling layers & 4 & 4\\
Number of fully-connected layers & 4 & 4\\
Fully-connected activation function & ReLU & ReLU\\ 

\hline\noalign{\smallskip}
\multicolumn{3}{c}{\emph{Optimization}} \\
\noalign{\smallskip}

Batch size & [32, 64] & 64 \\
Learning rate &  [$10^{-3}$--$10^{-7}$] & $10^{-6}$\\
Optimizer & [SGD, RMSProp, Adam] & Adam  \\
Loss function & \multicolumn{2}{c}{Categorical crossentropy}\\

\hline
\end{tabular}
\end{table*}

\subsubsection{Convolutional layers}
\label{sec:conv_layer}

The input for the convolutional layer is a tensor with the shape of the image height, width and depth.
When data is passing trough this layer it becomes abstracted by a feature map. 
During this step a filter matrix -- or kernel -- of a given size is convolved with parts of the image, by moving it with a given stride until the whole image is traversed. These layers can detect low-level features in the first steps, but can extract high-level features in later stages.
\subsubsection{Pooling layers}
\label{sec:pooling_layer}

The pooling layer is responsible for reducing the spatial size of the convolved image. It reduces the required computational power and it is also important for the extraction of dominant features of the image. 
We used max pooling in our model, which returns the maximum value  from each portion of an image: it selects important features as well as reduces noise.

\subsubsection{Fully-connected layers}
\label{sec:fully_conn_layer}

Fully-connected layers (also known as Dense layers) are responsible for the classification process as they can learn the non-linear combinations of the high-level features represented by the convolutional layers. As a final step, we use a softmax classification (basically a generalized version of a sigmoid function for multiple outputs), which classifies our images into separate classes of variable stars.

\subsubsection{Spatial Dropout}
\label{sec:spatial_dropout}

Data augmentation is crucial for a well-functioning deep neural network in the pre-process phase (see Section\,\ref{sec:data_aug}). However,  data augmentation alone is not always enough. One serious obstacle in applied machine learning is overfitting. A model is considered overfitted when it learned the features and their noise with high precision, but it poorly fits a new, unseen \report{data set}.  To be able to avoid it, one of the options is using Dropout layers. Dropout layers randomly neglect the output of a number of randomly selected neurons during training, in our case this was 20 percent. We used Spatial Dropout layers, which drop not just the nodes but the entire feature maps as well. These feature maps were not used by the next pooling process. Dropout layers offer a computationally cheap and remarkably effective regularization method to reduce overfitting and improve generalization error in deep neural networks. It helped us to be able to run the training much longer, so we could achieve very high accuracy.

\report{
\subsection{Optimizers and Learning rate} 
\label{sec:optimizers}
 We tested SGD (Stochastic Gradient Descent), RMSProp (Root Mean Square Propagation) and Adam (Adaptive Moment estimation) optimizers with various setups. After thorough testing we chose Adam  as the optimizer in our model. For the learning rate we tested various values between $10^{-3}$ and $10^{-7}$, in our model we chose a very low rate as $10^{-6}$.
}

\report{
\subsection{Early stopping} 
\label{sec:earlystopping}
We built in an EarlyStopping callback into the training method. This particular one is monitoring the change of the validation loss value, which is a key parameter by catching the signs of overfitting. In this case if the validation loss does not decrease by $10^{-4}$, the callback will run for another 7 additional epochs, stop the training process and save the best weight for further testing.
}

\subsection{Random forest classifier} 
\label{sec:random_forest}

\report{To compare our results with a method that only uses pre-computer features, we trained a Random Forest (RF) classifier \citep{RF} as well. RF is a machine learning algorithm that uses labeled (supervised) data and ensembles the results of several decision trees to classify the input into several classes. Here we use the RF that is implemented in the \texttt{scikit-learn} package \citep{scikit}. }

\report{The training set was created from the amplitude ($A$) and $R_{21} =A_2/A_1$, $\phi_{21} =\phi_2-2\phi_1$ Fourier-parameters of the original sample available in the OGLE-III database. The testing set is consist of the same parameters using both OGLE-III and OGLE-IV databases. As these values for eclipsing binaries are not present in the catalog, we calculate them utilising the periods from the OGLE database. To balance the number of samples in the five classes, we sampled dummy parameters from Gaussian distributions with means and standard deviations equal to the original parameters and $10^{-3}$ (for $A$), $10^{-4}$ (for $R_{21}$, $\phi_{21}$), respectively. The ratio of training and testing sample was $80\%-20\%$. }

\report{To get a robust, reproducible result and to prevent overfitting, we used 1000 trees in the "forest", set the maximum depth of each tree to 10, the minimum number of samples required to split an internal node to 10, the minimum number of feature per lead node to 5, and the random state to 40. }

\subsection{Evaluation Metrics} 
\label{sec:evaluation_metrics}

The performance of a trained machine-learning algorithm can be quantitatively characterised through several evaluation metrics. The one where the input and predicted class labels are plotted against each other is called a confusion matrix, where the predicted class is indicated in each column and the actual class in each row. This method allows us to visualise the number of true/false positives and negatives. In the best-case scenario, if the matrix is normalised to unity, we expect the confusion matrix to be purely diagonal, with
non-zero elements on the diagonal, and zero elements otherwise.

Precision is defined as:
\begin{equation}
    \mathrm{Precision} = \frac{TP}{TP + FP},
\end{equation}
where TP is the number of true positives and FP is the number of false positives. Precision shows that how precise the final model is out of those predicted positive, i.e. how many of predicted positives are actual positive.

Recall is defined as:
\begin{equation}
    \mathrm{Recall} = \frac{TP}{TP + FN},
\end{equation}
where TP is the number of true positives and FN is the number of false negatives. Recall shows that how many of the actual positives are labelled by the model as true positives.

From the last two metrics the F1-score can be calculated, which is the harmonic average of the precision and recall:
\begin{equation}
    F1 = 2 \cdot \frac{\mathrm{Precision} \cdot \mathrm{Recall}}{\mathrm{Precision }+ \mathrm{Recall}}.
\end{equation}
F1-score can measure the accuracy of the model, which returns a value between 0 and 1, where the latter corresponds to a better model.
   
   \begin{figure}
   \centering
   \includegraphics[width=\columnwidth]{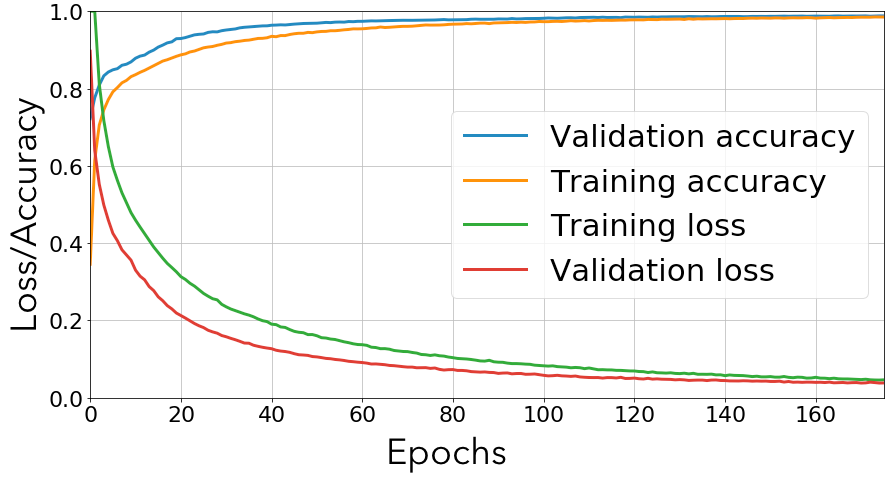}
      \caption{\report{Accuracy and loss of the training and validation process. Orange curve: training accuracy, red curve: validation accuracy. Blue curve: training loss, green curve: validation loss.}
              }
         \label{fig:train_loss_curves}
   \end{figure}

\section{Results}
\label{sec:results}

\subsection{Training and validation}
\label{sec:training_and_validation}

Our final \report{data set} contained 125\,000 images, 25\,000 from each type. This \report{data set} was subdivided into three different parts (70--15--15 $\%$), choosing images without any overlap for training, validation and testing purposes, respectively. 87\,500 images were used for training and 18\,750 for validation (see Table\,\ref{Tab:imageNumbers}). The process that goes through these two phases (training and validation) is called an epoch.
\report{We used GPU-accelerated computers provided by the MTA Cloud\footnote{\url{https://cloud.mta.hu}} and the Konkoly Observatory for this research.
Each training and validation epoch took about 290 seconds on a NVidia Tesla K80 GPU supported computer and 62 seconds with a NVidia GeForce RTX 2080 Ti GPU card.}
We were constantly checking the accuracy and loss values. \report{An EarlyStopping callback stopped the training process after 173 full epochs.} Inspecting the log files in TensorBoard showed no overfitting, after the 173$^{rd}$ epoch we reached 98.5\% training accuracy for the complete model (see Figure\,\ref{fig:train_loss_curves}).

 \begin{figure}
   \centering
   \includegraphics[width=\columnwidth]{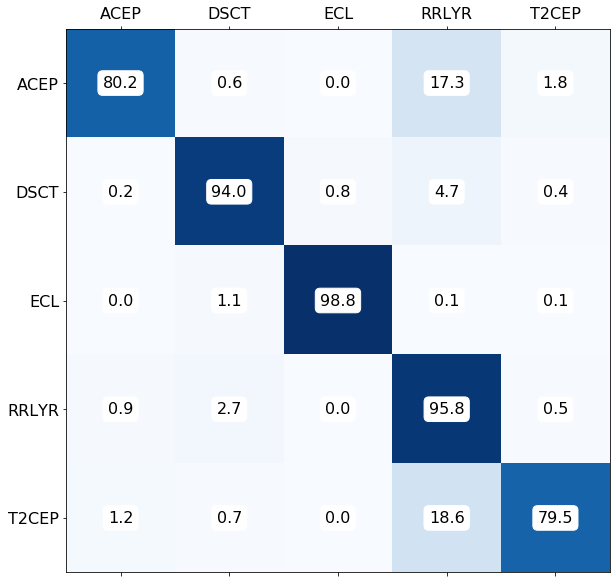}
      \caption{Test result on the OGLE-III data.}
         \label{fig:ogleIII_res}
   \end{figure}

   \begin{figure}
   \centering
   \includegraphics[width=\columnwidth]{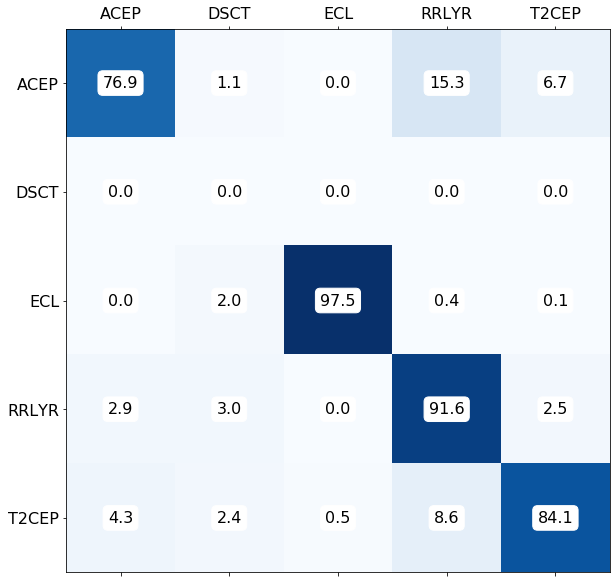}
      \caption{Test result on the OGLE-IV data.} 
         \label{fig:ogleIV_res}
   \end{figure}

\subsection{Testing the model}
\label{sec:testing}

We made two separate prediction tests on our model, the first one ran on the previously mentioned OGLE-III data.

As our original \report{data set} was divided randomly into three different parts, 87\,500 images were used for training and 18\,750 for validation, the remaining 18\,750 light curve images were for testing purposes. This test \report{data set} contained 3\,750 images from each variable star type and the test method ran through all light curves, using the weights from our trained model. We received a predicted label for each image and in the end of the test we could see how well our model is working with the OGLE-III LMC data (see Figure\,\ref{fig:ogleIII_res}).

For our second test we generated 10\,000 augmented samples (2\,500 from each type) from the OGLE-IV database (see Table\,\ref{Tab:imageNumbers}). The method was the same as before, we made predictions on each image, using the weights from the trained network (see Figure\,\ref{fig:ogleIV_res} and Table\,\ref{Tab:scoresIV}). 
\report{Comparing the two confusion matrices it is clearly visible that our trained model is working well and can classify variable stars from a different database.} 

Tables\,\ref{Tab:scoresIII} and \ref{Tab:scoresIV} show that a CNN trained on phase-folded light curves can classify variable stars with very high accuracy. Based on our results, we conclude that our model can efficiently distinguish between the eclipsing binaries and the periodic variables, like RR Lyrae stars. However, we note that due to similar light curve shapes and noise features, and due to the fact that we refrained from using the period as an input parameter 
we got false predictions for the pulsating stars in our second test. ACep, RRLyr and T2Cep stars were especially vulnerable to false prediction, while DSct stars were not available for OGLE-IV, as we mentioned before.

In this research we focused only on the light curve shapes, but it will be possible in the future to insert other data (most importantly, the period) into a more complex multi-channel network which could handle more inputs, image and numerical data as well.

\begin{table}
\caption{Classification report for the 5 classes in the OGLE-III
data-set. \report{Numbers correspond to the CNN and RF trainings, respectively.} The confusion matrix for this report is shown in Figure\,\ref{fig:ogleIII_res}. 
}
\label{Tab:scoresIII}
\begin{tabular}{rccc}
\hline
 & Precision & Recall & F1 score \\
 \hline
 \hline
ACep                 & 0.803 \report{/0.93}     & 0.972 \report{/0.95}  & 0.879 \report{/0.94}   \\
DSct                 & 0.939 \report{/0.87}    & 0.949 \report{/0.89} & 0.944 \report{/0.88}  \\
ECL                  & 0.987 \report{/0.98}   & 0.992 \report{/0.95} & 0.989 \report{/0.96} \\
RRLyr                & 0.959 \report{/0.88}   & 0.702 \report{/0.86} & 0.810 \report{/0.87} \\
T2Cep                & 0.795 \report{/0.96}   & 0.966 \report{/0.96} & 0.872 \report{/0.96} \\ \hline
Average               & 0.897 \report{/0.92}    & 0.916 \report{/0.92} & 0.899 \report{/0.92}\\
\hline
\end{tabular}
\end{table}

\begin{table}
\caption{Classification report for the 5 classes in the OGLE-IV
data-set. \report{Numbers correspond to the CNN and RF trainings, respectively.} The confusion matrix for this report is shown in Figure\,\ref{fig:ogleIV_res}.
}
\label{Tab:scoresIV}
\begin{tabular}{rccc}
\hline
 & Precision & Recall & F1 score \\
\hline
\hline
ACep                 & 0.769 \report{/0.89}    & 0.914 \report{/0.62} & 0.835 \report{/0.73}   \\
DSct                 & \nodata     & \nodata  & \nodata   \\
ECL                  & 0.975 \report{/0.96}   & 0.994 \report{/0.94} & 0.984 \report{/0.95}  \\
RRLyr                & 0.916 \report{/0.72}   & 0.790 \report{/0.90} & 0.849 \report{/0.80} \\
T2Cep                & 0.841 \report{/0.99}    & 0.900 \report{/0.75} & 0.870 \report{/0.85} \\ \hline
Average               & 0.875 \report{/0.89}    & 0.900 \report{/0.80} & 0.885 \report{/0.83} \\ 
\hline
\end{tabular}
\end{table}

\begin{table}
\caption{\report{Classification report for the 5 classes in the OGLE-III
data set of the RF method.}}
\label{Tab:scores_rf_III}
\begin{adjustwidth}{-2cm}{}
\begin{tabular}{rc|ccccc}
&   & \multicolumn{5}{c}{Predicted class}   \\
  
&   &   ACep &  DSct    &   ECL &   RRLyr   &   T2Cep   \\
\hline
\multirow{5}{*}{\rotatebox{90}{True class}}& ACep  &   95.46 &   0.08  &   0.00  &   4.46  &   0.00\\
& DSct  &   0.86  &   89.26 &   1.85  &   5.46  &   2.57\\
& ECL   &   0.00  &   3.91  &   95.11 &   0.14  &   0.85\\
& RRLyr &   4.50  &   7.85  &   0.42  &   86.45 &   0.77\\
& T2Cep &   2.12  &   0.89  &   0.00  &   1.46  &   95.53\\

\end{tabular}
\end{adjustwidth}
\end{table}

\begin{table}   
\caption{\report{Classification report for the 5 classes in the OGLE-IV
data set of the RF method.}}
\label{Tab:scores_rf_IV}
\begin{adjustwidth}{-2cm}{}
\begin{tabular}{rc|ccccc}
&   &   \multicolumn{5}{c}{Predicted class}   \\
  
&   &   ACep &  DSct    &   ECL &   RRLyr   &   T2Cep   \\
\hline
\multirow{5}{*}{\rotatebox{90}{True class}} & ACep  &   62.02 &   9.41 &   0.00  &   28.57 &   0.00\\
& DSct  &   \nodata  &   \nodata  &   \nodata  &   \nodata  &   \nodata\\
& ECL   &   0.00  &   4.92  &   94.06 &   0.14  &   0.89\\
& RRLyr &   4.62  &   5.16  &   0.07  &   89.92 &   0.24\\
& T2Cep &   3.03  &   11.67 &   4.20  &   6.52  &   74.57\\

\end{tabular}
\end{adjustwidth}
\end{table}

\begin{table}
\caption{The number of images used in the various steps.
}
\label{Tab:imageNumbers}
\begin{tabular}{rccc}
\hline
Survey & Training & Validation & Testing \\ \hline
\hline
OGLE-III    & 87\,500     & 18\,750  & 18\,750   \\
OGLE-IV     & \nodata   & \nodata  & 10\,000  \\ 
\hline
\end{tabular}
\end{table}

\section{\report{Discussion}}

\report{Our method uses a relatively new approach to classify the light curves of periodic variable stars, on the contrary of similar NNs, we do not use the time stamps of the measurements directly nor do we transform the light curves into another space. Instead, we look only at the light curve shape characteristics, which is achieved by phase-folding to increase the sampling within a cycle to be able to describe the shape more precisely. To compare our results with a more traditional method, we trained a RF algorithm using amplitudes and $R_{21}$, $\phi_{21}$ Fourier-parameters that best characterise the light curves shapes. Comparing Figure\,\ref{fig:ogleIII_res} to Table\,\ref{Tab:scoresIII}, and Figure\,\ref{fig:ogleIV_res} to Table\,\ref{Tab:scoresIV}, i.e. OGLE-III and OGLE-IV results, respectively, we can see that overall our CNN algorithm predicts better. Two cases where the CNN spectacularly preforms worse are the OGLE-III ACep and T2Cep classes, but as we do a transfer learning, i.e. test these methods on the independent OGLE-IV data set, we find that our CNN gives similar results as before, while RF gives worse results by 15--16\% for the mentioned classes.}

\report{A major point that we should discuss is the quality of training sets. As we described in Section\,\ref{sec:data}, we did not clear our sample, i.e. we included outliers and low-quality data, which makes the training more realistic and a harder task for the CNN to learn the weights. However, these bad values have a much subtle impact on the calculation of Fourier-parameters, making the RF result more boosted.}

\report{
Anomalous Cepheids are relatively larger mass (1--2 M$_\sun$) variable stars that lie in the classical instability strip. They follow a period-lumonisity relation, and their luminosity is between the classical and Type\,II Cepheids'. They are pulsating in the fundamental mode or the first overtone with a period shorter than 2 days. Their light curve is characterized by a steeper ascending branch which is followed by a shallower descending branch. Usually a bump is present at the bottom of the ascending branch. These features make it very hard or nearly impossible to distinguish them from RR Lyrae stars without known distances, i.e. their absolute brightness.}

\report{One of the main goals of our work is to see whether a CNN can distinguish Anomalous Cepheids from other variable stars based only on the light curve characteristics. From Figure\,\ref{fig:ogleIII_res} and Table\,\ref{Tab:scores_rf_III} we can see that our CNN was able to well-classify the 80.2\%, while the RF, which is based on pre-computed features, well-classified the 95.46\% of ACeps in the OGLE-III sample. As it is expected, the majority of the misclassifications are labeled as RR Lyrae stars (17.3\% and 4.5\%). These results show that there are hints of differences that make it possible to separate ACeps without known distances. Regarding our work, it is interesting that the CNN classifies ACeps about 15\% worse than the RF. However, if we test these methods using the independent OGLE-IV database (see Figure\,\ref{fig:ogleIV_res} and Table\,\ref{Tab:scoresIV}), our CNN still performs near 80\% (76.9\%), while the performance of RF drops down to 62\%. However, this decrease is not entirely surprising, as RFs are restricted to predict within the range of input parameters, i.e. they are not useful for transfer learning.}

\report{These results mean that image-based CNN classification may take place in applications where the training set slightly differs from the data set on which the prediction will be made.}

\section{Conclusions and Future Prospects}
\label{sec:conclusion}

In this work we trained a deep neural network to be able to distinguish different types of variable stars, based on light curve images generated from the OGLE-III database. To be able to do this, we generated a data-augmented image \report{data set}, containing equal amount of images from the chosen 5 types of variable stars. After thorough testing, a Convolutional Neural Network was created, which learned the different light curve features with high level of accuracy.

We demonstrated that image-based variable star classification  is a viable option using a Convolutional Neural Network. This type of machine learning method can learn both the high and low level features of a folded variable star light curve with high level of accuracy, in our case between 80 and 99 $\%$, based on OGLE-III data. 
It is clearly visible from our results that additional data (e.g. period) could increase the classification accuracy. We are working on a multi-channel network, where additional important parameters can be added as input, this way we expect that the classification accuracy of different variable star types will continue to increase. Our future plans also include generating light curve images for all variable stars in the OGLE-III LMC and SMC fields and using their subtypes available (e.g. RRab/RRc/RRd instead of RRLyr), as well. This way we would have a vast amount of training data and our model could be more specific and reliable.

\begin{table}
\caption{Approximate computational runtimes in minutes. \report{Numbers correspond to the NVidia Tesla K80, and NVIDIA RTX 2080 Ti GPU cards, respectively.}
}
\label{Tab:compTime}

\begin{tabular}{lccc}
\hline
Survey & Preprocess & Training & Testing \\ 
\hline
\hline
OGLE-III    & 265     & 167 /\report{61}  & 28 /\report{12}  \\
125\,000 images   \\
OGLE-IV     & 30   & \nodata  & 23 /\report{5}  \\
10\,000 images   \\ 
\hline
\end{tabular}
\end{table}

Training and testing a Convolutional Neural Network requires vast amount of computational time and capacity. We used GPU-accelerated computers in this research.
However, making predictions (i.e. classification itself) is possible with the saved weight file on any commercial computers. Predicting a label for one image takes just a fraction of a second (e.g. 0.13 seconds on a simple laptop), meaning that predictions even on large amount of light curve images can be made in a very short time (see Table\,\ref{Tab:compTime}).

A novel way of variable star classification would make a difference in the interpretation of the billions of light curves available today (and more to come). The Zwicky Transient Facility \citep[][ZTF]{Masci_2018} produced $\sim$1 billion light curves with more than 20 data points at different epochs and this number is continuously growing. The All-Sky Automated Survey for Supernovae \citep[][ASAS-SN]{Shappee_2014,Kochanek_2017} database contains 61.5 million light curves currently out of which 666\,500 objects were found to be variables. At least 62\,500 of them have unreliable classifications. The First Catalog of Variable Stars Measured by ATLAS \citep[][ATLAS-VAR]{Heinze_2018} detected 4.7 million variable objects already, but only 214\,000 of them received specific classifications. The Transiting Exoplanet Survey Satellite \citep[][TESS]{Ricker_2015} is in its second year of operations and keeps collecting excellent quality data from space with high cadence, like Gaia \citep{GaiaMission} does for billions of sources on the entire sky with lower cadence, and only a small fraction of them is classified accurately \citep[below 1\%, see e.g.][]{Molnar_2018,Gaia_2019,Marton_2019}. Future surveys, like the Rubin Observatory Legacy Survey of Space and Time \citep[][LSST]{Ivezic_2019} will further increase the number of objects with available time series data. One can see that astronomy needs accurate and efficient methods to rapidly analyse and classify variable objects on the sky. In the upcoming papers of the series we will explore further light curve data with the ultimate goal of providing such methods using image-based classification.

\acknowledgements
This project has been supported by the Lend\"ulet Program  of the Hungarian Academy of Sciences, project No. LP2018-7/2019, the NKFI KH-130526 and NKFI K-131508 grants, the Hungarian OTKA Grant No. 119993, and the MW-Gaia COST Action (CA 18104). GM was supported by the Hungarian National Research, Development and Innovation Office (NKFIH) grant PD-128360. GM acknowledges partial support from the EC Horizon 2020 project OPTICON (730890) and the ESA PRODEX contract nr. 4000129910.
This project has received funding from the European Research Council (ERC) under the European Union's Horizon 2020 research and innovation programme under grant agreement No 716155 (SACCRED).
On behalf of \textit{"Analysis of space-borne photometric data"} project we thank for the usage of MTA Cloud (\url{https://cloud.mta.hu}) that significantly helped us achieving the results published in this paper.

\software{Python \citep{numpy},
Numpy \citep{numpy},
Pandas \citep{pandas},
Scikit-learn \citep{scikit},
Tensorflow \citep{Tensorflow},
Keras \citep{Keras} }

\bibliography{references}{}
\bibliographystyle{aasjournal}



\end{document}